# Built-in Homojunction Dominated Intrinsically Rectifying-Resistive Switching in NiO Nanodots for Selection Device-Free Memory Application

By *Zhong Sun*, *Linlin Wei*, *Ce Feng*, *Peixian Miao*, *Meiqi Guo*, *Huaixin Yang*, *Jianqi Li*, and *Yonggang Zhao*[*]

[*]     Z. Sun, C. Feng, Dr. P. Miao, M. Guo, Prof. Y. Zhao
Department of Physics and State Key Laboratory of Low-Dimensional Quantum Physics,
Tsinghua University
Beijing 100084, China
E-mail: ygzhao@tsinghua.edu.cn
        L. Wei, Prof. H. Yang, Prof. J. Li
Beijing National Laboratory for Condensed Matter Physics, Chinese Academy of Sciences
Beijing 100190, China
        Z. Sun, C. Feng, Dr. P. Miao, M. Guo, Prof. J. Li, Prof. Y. Zhao
Collaborative Innovation Center of Quantum Matter
Beijing 100084, China

The intrinsically rectifying-resistive switching (IR-RS) has been regarded as an effective way to address the crosstalk issue, due to the Schottky diodes formed at the metal/oxide interfaces in the ON states to suppress the sneak current at reverse biases. In this letter, we report for the first time another type of IR-RS that is related to the built-in homojunction. The IR-RS study was usually limited to macroscopic samples with micron-order pad-type electrodes, while this work is on NiO nanodots fabricated with ultrathin anodic-aluminum-oxide templates and acting as nanoscaled analogs of real devices. The NiO nanodots show high storage density and high uniformity, and the IR-RS behaviors are of good device performances in terms of retention, endurance, switching ratio and rectification ratio. The feasibility of the IR-RS for selection device-free memory application has been demonstrated, by calculating the maximum crossbar array size under the worst-case scenario to be 3 Mbit.

## 1. Introduction

Memristive device utilized as resistive random access memory (RRAM) has been considered as the most promising next-generation nonvolatile memory, due to its simple structure, high scalability and outstanding performance.[1,2] It can be applied to the crossbar structure to achieve the highest storage density with the smallest cell size of $4F^2$ ($F$ is the minimum feature size). In this case, however, it will suffer from the so-called crosstalk issue,[3] which refers to the interference between the read current through the selected cell and the sneak current through the unselected cells. To overcome this obstacle, the integration of selection devices such as transistors,[3,4] diodes,[5,6] and selectors[7,8] has been suggested. However, this approach is not favorable to device scaling, and the integration process is complex.[9,10] Another solution called complementary resistive switch (CRS) has also been proposed,[11,12] but the inherently destructive read operation has severely limited its application. Recently, the intrinsically rectifying-resistive switching (IR-RS) has been found as an effective way to address the crosstalk issue.[9,10,13-21] In these systems, Schottky diodes are formed at one of the metal/oxide interfaces in the ON states to suppress the sneak current at reverse biases, and the diodes are annihilated,[9,10,13,16-18] or switched in some cases,[19-21] to reversed ones at the counter-interfaces in the OFF states through oxygen migration. Basically, the IR-RS is a type of bipolar resistive switching (RS). Besides the interface-dominated IR-RS, the IR-RS originating from the built-in isotype/anisotype homojunction is also possible in ohmic metal/oxide/metal structures, since the concentration gradient of charge carriers caused by the ion migration under electric field has been demonstrated.[22-24] Generally, a built-in homojunction should allow lower reverse leakage current than a Schottky diode,[25] which is more beneficial to addressing the crosstalk issue. However, the built-in homojunction dominated IR-RS behaviors have not been reported so far. In contrast to most memristive metal oxides of *n*-type conductivity, NiO as a prototypical *p*-type oxide is prone to form ohmic contacts with the frequently-used metal electrodes such as Pt and Au.[26] Previous work have demonstrated that the local carrier concentration in NiO can be modulated by oxygen migration due to the cooperative impact of Ni vacancies and oxygen vacancies, resulting in the inhomogeneous distribution of hole carriers.[27,28] This favors the observation of built-in homojunction dominated IR-RS in an ohmic metal/NiO/metal structure.

On the other hand, IR-RS in the previous work were limited to the macroscopic samples with micron-order pad-type electrodes.[9,10,13,15-20] To check the validity of device application, it is imperative to reduce the dimension to the real device size. In the case of NiO, in fact, the nanodevice is conducive to the observation of IR-RS, due to the dominance of the bipolar electric-field-induced oxygen migration over the unipolar thermal effect in nanoscale NiO.[29] The bottom-up self-assembly with nanoporous templates has been demonstrated as a powerful approach to fabricate nanostructures, with advantages over the lithographic techniques, including large scale, low cost, short processing period, and sublithographic scalability.[30-32] Son *et al.* has studied the RS behaviors of NiO nanocapacitors, which were fabricated on a graphene/Nb:SrTiO$_3$ substrate with an anodic aluminum oxide (AAO) template.[31] However, they just observed the ordinary unipolar RS, which may be attributed to their specific device configuration.

In this work, we have fabricated NiO nanodots of high storage density and high uniformity, by utilizing AAO templates. The NiO nanodots show two types of IR-RS behaviors, namely self-rectifying RS and switchable diode-like RS, with good device performances in terms of retention, endurance, switching ratio and rectification ratio. Due to the *p*-type nature of NiO and the respective work functions of NiO and metal electrodes, the contributions of both metal/oxide interfaces are ruled out, and the IR-RS mechanism is attributed to the built-in isotype homojunction that is modulated by oxygen migration. This is the first work reporting the IR-RS dominated by the built-in homojunction, differing from the previous cases dominated by the interfacial Schottky diodes. Additionally, the IR-RS was observed on nanoscaled analogs of real devices, and its feasibility for selection device-free memory application has been demonstrated by calculating the corresponding maximum crossbar array size.

## 2. Results and discussion

Ultrathin AAO template with a nanopore depth of 200 nm was employed to fabricate NiO nanodots by using pulsed laser deposition (PLD), as depicted in **Figure 1**a. After removal of the AAO template, well-ordered NiO nanodots attached to the susbtrate were obtained. The atomic force microscopy (AFM) image in Figure 1b indicates that, the distribution of NiO nanodots is very uniform on a large scale of $10\times10$ μm$^2$. The spacing of the nanodots is around 100 nm, which is consistent with the AAO interpore distance (Figure S1, Supporting Information), and guarantees a high storage density of 110 Gb inch$^{-2}$. The distribution of cumulative probability in Figure 1c reveals the uniformity of the fabricated NiO nanodots quantitatively. Calculated from the AFM height profile, the average height is determined to be $23.1\pm2.6$ nm (coefficient of variation ~ 11%). To avoid the influence of convolution with AFM tip, the diameters of NiO nanodots were determined to be $45.6\pm4.1$ nm (coefficient of variation ~ 9%) by using scanning electron microscopy (SEM). The transmission electron microscopy (TEM) result in Figure 1d shows the trapezoid-like shape of NiO nanodots, which was caused by the progressive narrowing of nanopores during deposition. The high-resolution TEM (HRTEM) image of an individual NiO nanodot is shown in Figure 1e, where nanocrystallites with various orientations can be recognized. The inset of Figure 1e shows the corresponding fast Fourier transform (FFT) analysis with the NiO (111) and (200) planes labeled. The polycrystalline nature of NiO nanodots is beneficial to the occurrence of RS, owing to the facilitated ion migration along the grain boundaries.[33]

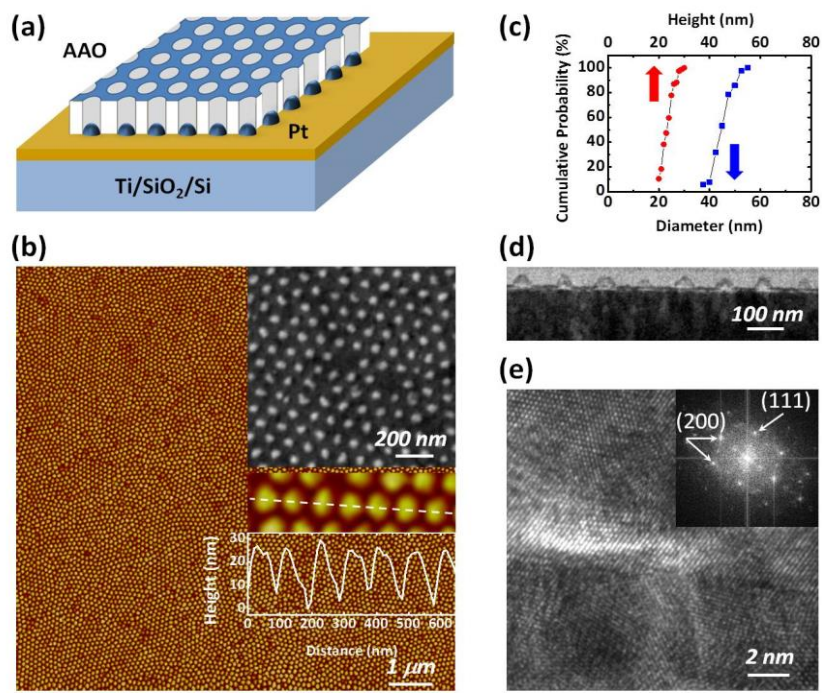

**Figure 1.** NiO nanodots prepared with AAO templates. a) Schematic of NiO nanodots grown in the pores of an AAO template via pulsed laser deposition. b) AFM morphology of fabricated NiO nanodots. The insets show the SEM image and the AFM height profile of NiO nanodots. c) Probability distributions of height and lateral diameter of NiO nanodots. d) Cross-sectional TEM image of an array of NiO nanodots. e) HRTEM image of an individual NiO nanodot. The inset is the corresponding FFT analysis.

The RS behaviors of NiO nanodots were measured in ambient conditions with a conductive AFM (CAFM), which was equipped with a Pt/Ir-coated tip whose curvature radius is 20 nm. The CAFM can detect tiny current as small as 1 pA. During the measurement, the bias voltage was applied to the Pt bottom electrode (BE) with the tip grounded (**Figure 2**a). To measure the current-voltage (*I-V*) characteristics of NiO nanodots, the tip was located on a single nanodot, and the voltage was swept in a sequence of -10 V → 10 V → -10 V, which is the fixed voltage-sweep scheme of the instrument with a compliance current of 12.28 nA. The results indicate that, a single NiO nanodot shows various *I-V* characteristics of bipolar RS (Figure 2b-d), and the corresponding *I-V* curves in the semi-log scale are shown in the Supporting Information (Figure S2). Due to the variety of the bipolar RS behaviors observed here, and their difference from the normal bipolar RS, the state with higher (lower) currents at positive voltages is defined as ON (OFF) state. For the normal bipolar RS, the *I-V* curve of the ON state is usually symmetric,[34] and thus the currents at negative voltages in the ON state are also higher than those in the OFF state. However, for the bipolar RS of NiO nanodots, the *I-V* curves of the ON states are of rectifying characteristics, resulting in that the currents in the ON state may be lower than those in the OFF state at negative voltages, as shown in Figure 2c-d. The bipolar RS with rectifying characteristics is termed intrinsically rectifying-resistive switching (IR-RS). Though the NiO nanodots show various RS behaviors, these behaviors are all IR-RS, and can be used to solve the crosstalk problem in the RRAM applications. The IR-RS behaviors of NiO nanodots can be classified into two types, namely self-rectifying RS and switchable diode-like RS, based on the difference of *I-V* characteristics in the OFF state. We define a diode with forward direction at positive (negative) voltages as positive-forward (negative-forward) diode. Specifically, in the self-rectifying RS, the nanodot is switched to the ON state with positive-forward diode-like behavior for positive voltages, and back to the OFF state with symmetric *I-V* characteristics for negative voltages, with the *I-V* curves of both states crossing at 0 V (Figure 2b). If the currents at negative voltages in the ON state are lower than those in the OFF state, the self-rectifying RS would be *non-crossing* at the origin of *I-V* plot (Figure 2c). In Figure 2b, due to the specific voltage-sweep scheme of our CAFM, the *I-V* hysteresis is not necessarily closed at -10 V, which is both the beginning and ending point of the sweep cycle, but the following sweep cycle indicates unambiguously that the NiO nanodot has been switched to the OFF state at negative voltages (Figure S3, Supporting Information). In the switchable diode-like RS, the positive and negative voltage-sweeps induce the ON state with positive-forward diode-like behavior and the OFF state with negative-forward diode-like behavior, respectively (Figure 2d). The switchable diode-like RS results from the excess ion migration at negative voltages, evidenced by the controlled experiment in which the voltage-sweep to -8 V results in the self-rectifying RS, while the voltage-sweep to -10 V leads to the switchable diode-like RS (Figure S4, Supporting Information). The occurrence of two types of IR-RS behaviors in the same voltage range is attributed to the randomness of ion migration on the nanoscale. In spite of the behavior difference at negative voltages, the behaviors at positive voltages are pretty uniform (Figure S5, Supporting Information). We have checked the IR-RS behavior of NiO nanodots with a different voltage-sweep scheme, in which a voltage-sweep from 0 V to a large voltage was applied to trigger the ON state, and a voltage-sweep in a smaller range was utilized to measure the static rectifying *I-V* characteristics of the ON state (Figure S6, Supporting Information), by employing an external source meter combined with the scanning probe microscope (SPM) setup. The small currents in both OFF- and ON-states are ascribed to the tiny contact radius between the CAFM tip and the sample, which was calculated to be 1.38 nm in our previous work.[28] If the sample was fabricated into crossbar with feature size of decades-of-nanometers, the working currents would be in the range of nanoamps to microamps, which would be more proper for practical applications.

Besides the *I-V* characteristics, the bipolar RS behaviors of NiO nanodots have also been demonstrated by CAFM mapping. In Figure 2e, the area in the yellow dashed box was written with a 5 V bias, and the entire area was read with a 2 V bias, showing that some nanodots have been changed to the ON state. A larger writing voltage of 8 V can switch nearly all the NiO nanodots to the ON state, but artifacts may be introduced in the morphology image due to the higher applied electrical field (Figure S7, Supporting Information). Unlike the bipolar RS of electrochemical metallization mechanism (ECM), the bipolar RS of valence change mechanism (VCM) does not need a threshold voltage for its occurrence, and a larger voltage or a longer biasing duration just facilitates the bipolar RS through enhanced anion migration.[35] Since a fresh tip can contact well with the Pt-BE and read out the current (Figure S8, Supporting Information), the absence of current in between the NiO nanodots in Figure 2e should be attributed to the morphologic degradation of CAFM tip caused by the strong electrical stress during the writing process. The ON states of NiO nanodots can be erased by applying a -5 V bias, resulting in the CAFM mapping with only current arising from some interstices between the nanodots (Figure 2f). In Figure 2g-j, we focus on the current mapping of a single nanodot. The nanodot was initially insulating with surrounding current arising from the naked Pt-BE (Figure 2h). After the writing operation with a 5 V bias, the nanodot was switched into the ON state with pronounced conductivity localized around the edge area of the nanodot (Figure 2i). The written ON-state is long-lived and can persist for at least 18 hours (Figure 2j), showing better performance than the NiO films investigated by CAFM mapping.[36] Both the *I-V* measurements and the CAFM mapping results demonstrate the stable IR-RS behaviors of NiO nanodots, which is useful for selection device-free memory application.

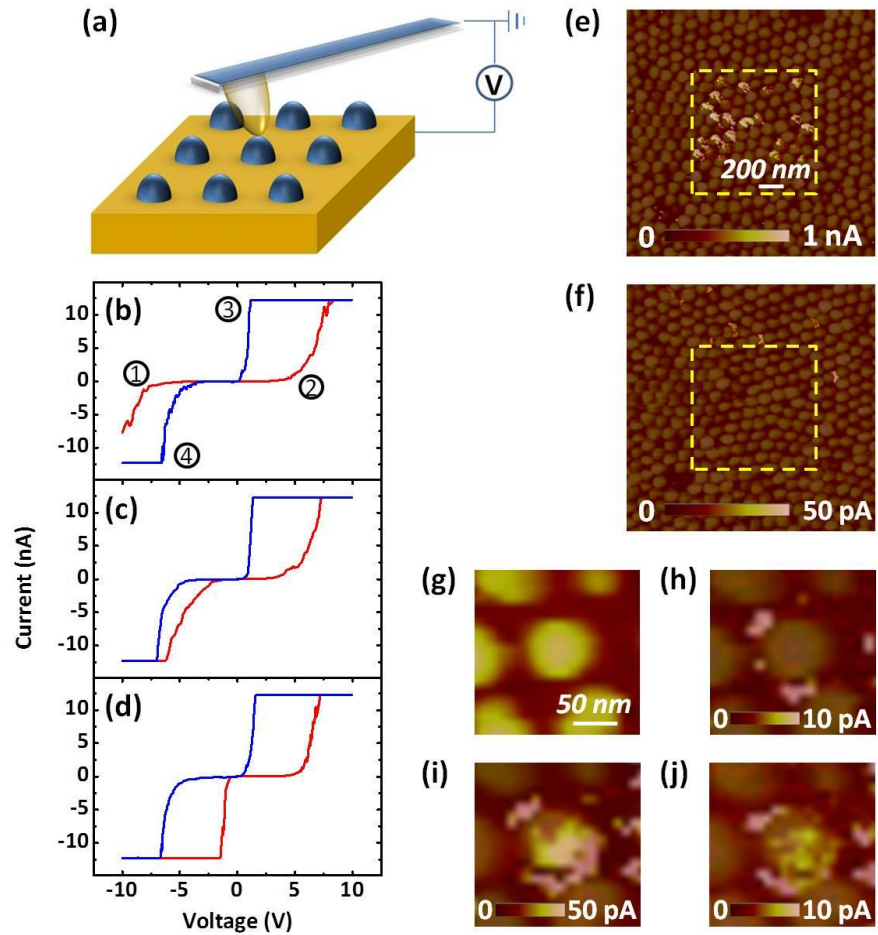

**Figure 2.** IR-RS of NiO nanodots. a) Schematic of CAFM measurement setup for NiO nanodots. b) Self-rectifying RS, c) *Non-crossing* self-rectifying RS, and d) Switchable diode-like RS of NiO nanodots. The red *I-V* curves were got by sweeping from -10 V to 10 V, and the blue curves were got by sweeping from 10 V to -10 V. The specific voltage-sweep sequence is labelled in b) as an example. Reading current mapping of e) written ON-state and f) erased OFF-state of NiO nanodots. g) AFM image containing one complete NiO nanodot.

h) Initial reading current mapping, i) Current mapping of the as-written ON-state, and j) Current mapping of the ON state after 18 hours superposed on the AFM topography. The writing voltage is 5 V, the erasing voltage is -5 V, and the reading voltage is 2 V.

As shown in **Figure 3**a, the endurance test indicates that the RS behavior of a single NiO nanodot can be cycled for nearly 900 times. In contrast, the endurance of other devices investigated with CAFM is usually limited to several tens cycles, which was ascribed to the drift of AFM tip.[37] Therefore, NiO nanodots are demonstrated with good device performance for memory applications. In Figure 3a, the currents in the ON state were extracted from the *I-V* curves at 2 V to achieve larger ratios between the OFF-state resistances and the ON-state resistances ($R_{OFF}/R_{ON}$), and a typical value of $10^4$ can be recognized. The $R_{OFF}/R_{ON}$ ratios with currents extracted at 1 V are around $10^3$ (Figure S9, Supporting Information). Actually, the $R_{OFF}/R_{ON}$ ratios at 2 V should be even larger if the compliance current is set higher and/or the current detection of the CAFM is more sensitive. Nevertheless, the $R_{OFF}/R_{ON}$ ratio of the IR-RS of NiO nanodots is a satisfactory result to facilitate the RRAM applications, since the value of the normal bipolar RS based on anion migration is usually less than $10^3$.[38] In Figure 3b, the rectification ratios of ON states of a NiO nanodot were calculated with the currents at $\pm$ 2 V, and the results are concentrated around $10^3$-$10^4$, with 72% larger than $10^3$. The results of $R_{OFF}/R_{ON}$ ratios and rectification ratios for 10 nanodots are shown in Figure 3c. The parameter dispersion of different NiO nanodots may be attributed to their differences in crystallinity and defect density. The average $R_{OFF}/R_{ON}$ ratio and rectification ratio for different NiO nanodots is calculated to be 2918 and 1572, respectively. Evaluated under the worst-case scenario with the $V_r$ bias scheme,[39] the large $R_{OFF}/R_{ON}$ ratios and rectification ratios observed in the IR-RS of NiO nanodots will in principle benefit the mitigation of the crosstalk issue. Therefore, a large crossbar array size can be expected in the subsequent calculation section. The nonlinear *I-V* characteristics of RS can also be exploited to address the crosstalk issue,[40] but the nonlinearity of the IR-RS of NiO nanodots is not that outstanding (Figure S10, Supporting Information), so the prominent rectifying characteristics would contribute most to alleviating the crosstalk problem in this work.

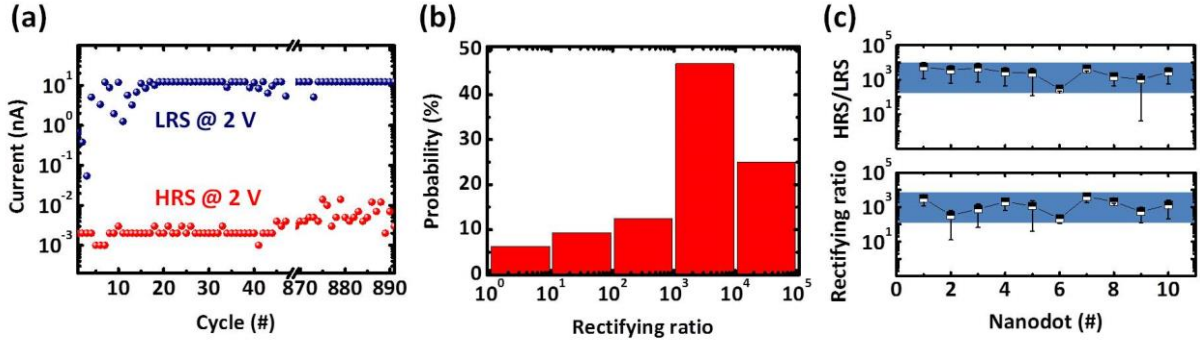

**Figure 3.** a) Endurance of the IR-RS tested on a NiO nanodot. b) Rectification ratio distribution of 32 ON-state curves for a single NiO nanodot. c) $R_{OFF}/R_{ON}$ ratios and rectification ratios of ten NiO nanodots, averaged from five consecutive *I-V* curves for each nanodot.

To investigate the mechanism of IR-RS, especially the role of oxygen playing in the IR-RS of NiO nanodots, we measured the RS behaviors in air and in vacuum, respectively. The NiO nanodot measured in air in the voltage range of -3 ↔ 3 V shows bipolar RS behavior (Figure 4a), despite the weak rectifying behavior that is due to the insufficient ion migration limited by the relatively low bias voltage for the VCM-dominated bipolar RS. However, the measurement in vacuum in the same voltage range doesn't show any RS behaviors at all with null current, unless the range was expanded to -10 ↔ 10 V (Figure 4b). The CAFM mapping shows that the NiO nanodots can be switched by a 10 V bias, with the conducting areas locating only at the edges of the nanodots. These observations indicate that oxygen plays an important role in the IR-RS of NiO nanodots. Yoshida *et al*.[41] demonstrated that the bipolar RS of NiO films is caused by oxygen migration, by using $^{18}$O tracer gas and time-of-flight secondary ion mass spectrometry (TOF-SIMS). In contrast, Lee *et al*.[42] suggested that it is caused mainly by the electrochemical redox action at the electrode-film interface instead, based on the observations by CAFM and Kelvin probe microscopy (KPM). Note that the descriptions of the two mechanisms capture only the main features of them for simplicity, but the oxygen migration mechanism will also inevitably result in the oxidation/reduction of cations in NiO. We have studied carefully the possible mechanisms for the IR-RS of NiO nanodots, leading to a conclusion that the mechanism involves the built-in homojunction induced by oxygen migration, while both NiO/metal interfaces should be ohmic during the RS events. Details are as follows.

Considering the *p*-type conductivity of NiO and work functions of NiO and Pt, it can be deduced that NiO/Pt-BE interface should not be involved in the RS of NiO nanodots. Because of the *p*-type nature of NiO, the NiO/Pt-BE interface cannot form a positive-forward Schottky diode when a bias voltage is applied to the Pt-BE. The energy band analysis on the contact between NiO and metal electrodes is available in the Supporting Information (Figure S11). The work function of NiO was reported to range from 3.8 eV to 5.4 eV,[43] and that of Pt is 5.7 eV,[44] so the contact between NiO and Pt has always been considered to be ohmic,[26] as it should be in this study.

If the electrochemical redox reaction dominates the IR-RS of NiO nanodots, the tip/NiO interface should be recognized as the switching region, and the RS processes can be described as the modulation of interfacial Schottky barrier by the oxygen incorporation/extraction. In this case, the RS behaviors should not be anticipated in a device capped with top electrode due to the blocking for oxygen exchange. However, in a Au/NiO nanodot/Pt structure, we have also observed the identical IR-RS behaviors (Figure S12, Supporting Information), which suggests that the electrochemical redox reaction should not play a significant role in the IR-RS of NiO nanodots. Moreover, the electrochemical redox mechanism cannot explain consistently both the self-rectifying RS and the switchable diode-like RS. Consequently we need to consider the possible mechanism related to oxygen migration in NiO.

We would like to point out that, despite the well-known defect chemistry of NiO from earlier studies[45-47] suggesting that the main defect species in NiO are Ni vacancies, which are responsible for the *p*-type conductivity of NiO, the oxygen vacancies are also expected to play an important role at low work-temperatures and low oxygen-pressures.[48] The NiO nanodots in this work were prepared at 500 ℃ and 1.5 Pa oxygen-pressure, which are almost the same with the conditions for NiO films in our previous work,[28] in which the oxygen vacancies in NiO have been observed directly by scanning transmission electron microscopy (STEM). Therefore, the oxygen vacancies should be involved in the IR-RS mechanism of NiO nanodots. Recently, several experimental work have demonstrated the oxygen migration in the bipolar RS of NiO.[41,49] Therefore, we will consider the oxygen migration instead of nickel migration for the IR-RS mechanism of NiO nanodots.

If the IR-RS of NiO nanodots is caused by oxygen migration, the tip/NiO interface should not be regarded as the switching region, because the downward oxygen migration at positive voltages will reduce the work function of NiO (Ref. 50) near the surface to facilitate the ohmic conduction at the tip/NiO interface, resulting in symmetric *I-V* characteristics instead of diode-like behavior in the ON state. Therefore, regardless of the specific mechanism, the tip/NiO interface cannot be the switching region of NiO nanodots, and it is indeed ohmic contact as evidenced by the symmetric *I-V* characteristics in the pristine state (Figure 4c). The larger work function (5.5 eV) of the Pt/Ir coating of CAFM tip[51] also supports the ohmic contact at the tip/NiO interface.

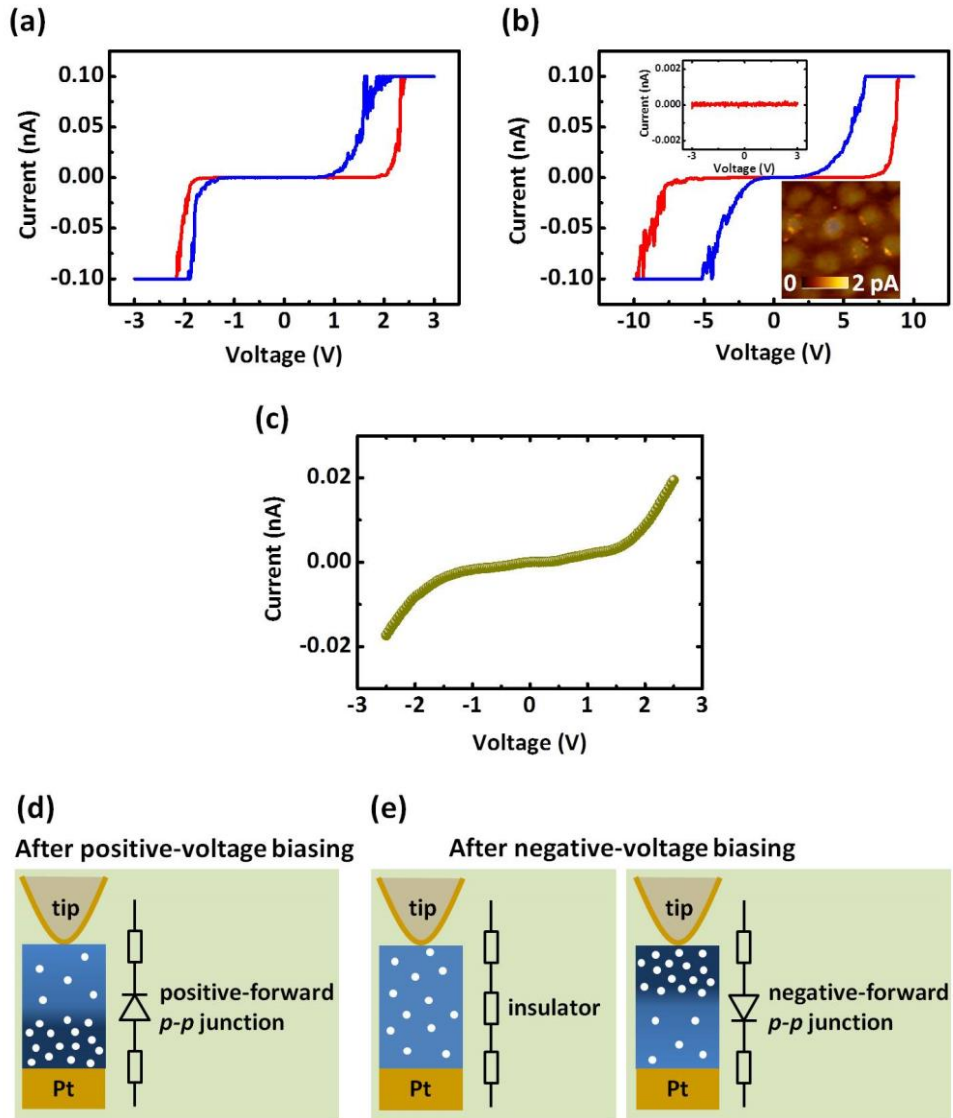

**Figure 4.** Mechanism of the IR-RS of NiO nanodots. RS behaviors of NiO nanodots measured a) in air and b) in vacuum with a compliance current of 100 pA. The upper left inset in b) shows the current measured in the voltage range from -3 V to 3 V. The bottom right inset in b) shows the reading current mapping of the written ON-states of NiO nanodots, where the writing voltage is 10 V, and the reading voltage is 3 V. c) Symmetric *I-V* characteristics of tip/NiO/Pt structure in the pristine state. Schematics of d) the positive-forward diode-like ON-state for both self-rectifying RS and switchable diode-like RS, and e) the insulating OFF-state for self-rectifying RS, and the negative-forward diode-like OFF-state for switchable diode-like RS in the built-in homojunction dominated mechanism. The tip/NiO and NiO/Pt-BE interfaces are both ohmic contacts during the RS events. The white circles represent oxygen ions.

Interestingly, the possible built-in homojunction formed by oxygen migration can be fully responsible for the results. The oxygen migration will modulate the local carrier concentration, resulting in a concentration gradient of charge carriers along the elelctric-field direction in the device, which gives rise to the rectifying characteristics of built-in isotype/anisotype homojunctions. In the case of NiO nanodots, the downward migration of oxygen ions at positive voltages will increase the hole carrier concentration in the region near the Pt-BE, due to the increased concentration of isolated Ni vacancies.[27] While in the region near the NiO surface, the hole carrier concentration is reduced due to the introduction of more oxygen vacancies. The concentration gradient of hole carriers in the NiO nanodot will result in a *p-p* isotype homojunction at equilibrium (Figure 4d), which can account for the positive-forward diode-like ON-state of NiO nanodots. It is worth mentioning that the formed homojunction can also be understood in the framework of work function change related to oxygen content. Though the conduction transition from *p*-type to *n*-type with decreasing oxygen content has been demonstrated in NiO,[52] it is not anticipated in this study, because a negative-forward diode will form between the *n*-type NiO near the surface and the tip coating in this case. Accompanying the oxygen migration, oxygen is incorporated from the ambient through the electrochemical redox reaction at the surface, which is enhanced by the large surface area of NiO nanodots. The consequent oxygen diffusion into the NiO nanodots will increase the overall concentration of hole carriers. At negative voltages, the oxygen ions drift backwards with oxygen extraction occuring at the surface, which results in the annihilation of built-in *p-p* homojunction, switching NiO nanodots to the insulating OFF-state (left panel in Figure 4e). In some cases, if the upward oxygen migration is excess or the outward oxygen extraction is small, the formed isotype homojunction may be reversed to a negative-forward *p-p* homojunction after the negative-voltage biasing (right panle in Figure 4e), resulting in the switchable diode-like RS behaviors. The schematics of respective band diagrams for Figure 4d-f are available in the Supporting Information (Figure S13). Therefore, it can be concluded that the mechanism of the IR-RS of NiO nanodots should be attributed to the built-in homojunction, differing from the previous cases which are dominated by the interfacial Schottky barriers.[9,10,13-21] It should be mentioned that, the moisture has been revealed to play an important and even deterministic role in some ECM- and VCM-based memories,[53-55] however, it should not be invoked in this work, since we have demonstrated, in our previous work, that the moisture does not have a notable influence on the CAFM-investigated RS behaviors of NiO.[28]

It is worth mentioning that the built-in homojunction characteristics caused by oxygen migration was reported in studying the electrical degradation of $SrTiO_3$ in the early '90s.[56] It has also been proposed that *p-n* anisotype homojunction was formed in $Bi_{0.8}Ca_{0.1}FeO_{3-\delta}$ in the $Bi_{0.8}Ca_{0.1}FeO_{3-\delta}/SrRuO_3$ heterostructure.[57] However, the presence of ferroelectricity and the related polarization-direction-controlled interface barrier[58,59] and domain wall conductivity[60] makes the issue complicated. Recently, a theoretical work showed that a *p-n* anisotype homojunction should form in the memristive devices through the migration of *n*-type dopants, with interfaces assumed to be purely ohmic.[24] However, the characteristics of built-in homojunction have always been missed in the RS of metal oxides, likely due to the dominant role of the interfacial barriers in the metal/oxide/metal structures, since most memristive oxides of interest are *n*-type, such as $SrTiO_3$,[61] $TiO_2$,[34] $ZnO$,[62] $HfO_2$,[63] and $ZrO_2$,[64] which are apt to form Schottky barriers with the metal electrodes, masking the effect of the built-in homojunctions.

To evaluate the validity of the IR-RS of NiO nanodots for addressing crosstalk issue, we have calculated the maximum crossbar array size with the one bitline pulled up (OBPU) approach,[65] in which a pull-up voltage ($V_{pu}$) and a pull-up resistance ($R_{pu}$) should be inputted (**Figure 5**a). We have assumed the worst-case scenario, where all unselected cells are in the ON-state (OFF-state) when reading a selected OFF-state (ON-state) cell.[65] With the increasing size of crossbar array, the difference between the output voltages ($V_{out}$) of ON-state and OFF-state will be reduced, due to the increase of sneak current paths. The read margin [$\Delta V_{out}/V_{pu} = (V_{out,OFF} - V_{out,ON})/V_{pu} \times 100\%$] of 10% has been frequently employed as the minimum creterion to calculate the maximum crossbar array size,[20] which is determined by the specific *I-V* hystersis loop characteristics. Giving a combination of ($V_{pu}$, $R_{pu}$), a maximum number $N_{max}$ of word/bit lines will be obtained. Since the *non-crossing* self-rectifying RS allows lower current at reverse biases, it is utilized for calculation to achieve the largest carossbar array size. The numerical simulation of the *I-V* characteristics of *non-crossing* self-rectifying RS is available in the Supporting Information (Figure S14). Figure 5b shows the results with a $V_{pu}$ range of 3.4 - 4.1 V, and an $R_{pu}$ range of 24000 – 31000 Ω. In the whole range, the calculated maximum number $N_{max}$ is larger than 1000, which guarantees a wide selection range of reading voltages. A maximum number of $N_{max}$=1701 is achieved with the combination of (26500 Ω, 4.1 V), leading to the maximum crossbar array size of 3 Mbit for selection device-free memory application, which shows obvious superiority over most cases of interface-dominated IR-RS.[13,17] Figure 5c shows the read margin evolution with the number of word/bit lines for a given ($V_{pu}$, $R_{pu}$). The detailed calculations are available in the Supporting Information (Figure S15). Therefore, NiO is suggested as a promising material for memory applications, especially regarding its simple constituent and compatibility with complementary metal-oxide-semiconductor (CMOS) process.[41,66] Though the built-in homojunction dominated IR-RS in this work is achieved in the nanodot form as a conceptual demonstration, we believe it should also apply to the crossbar structure, since it relies only on the ohmic contacts and the generation of concentration gradient of charge carriers. Further experimental realization will be appreciated.

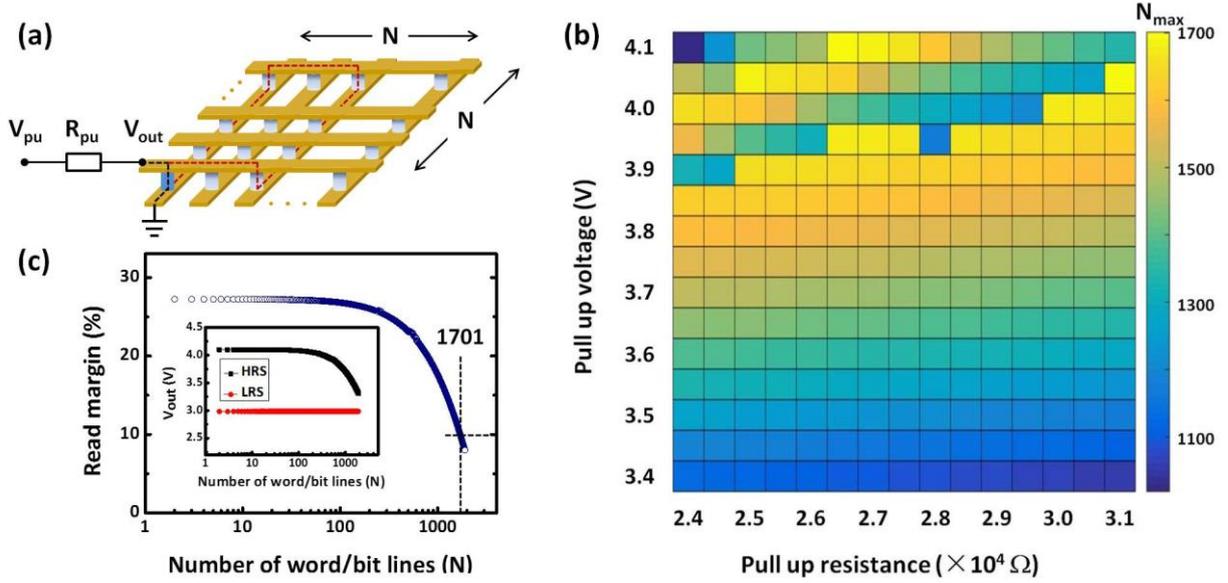

**Figure 5.** Calculation of the maximum crossbar array size for the self-rectifying RS of NiO nanodots. a) Reading a OFF-state cell under the worst-case scenario with the OBPU approach, where all unselected cells are in the positive-forward diode-like ON-state. The schematic for reading a positive-forward diode-like ON-state cell is not shown, where all unselected cell are in OFF-state. $V_{out}$ is the readout voltage. The black dash line indicates the read current path, and the red one indicates a representative sneak current path. b) Calculated maximum number $N_{max}$ of word/bit lines for different combinations of ($R_{pu}$, $V_{pu}$), with $R_{pu}$ in the range of 24000-31000 Ω, and $V_{pu}$ in the range of 3.4-4.1 V. A maximum $N_{max}$ of 1701 is achieved with combination of (26500 Ω, 4.1 V). c) Read margin evolution with the number of word/bit lines for the combination of (26500 Ω, 4.1 V). Inset shows the readout voltage evolutions of both ON-state and OFF-state.

## 3. Conclusion

In summary, we have fabricated NiO nanodots of high density and uniformity with ultrathin AAO templates. Investigated by CAFM, NiO nanodots show typical IR-RS behaviors, including self-rectifying RS and switchable diode-like RS, and the good device performance in terms of retention, endurance, switching ratio and rectification ratio have been indicated. According to the *p*-type nature of NiO and the respective work functions of NiO and metal electrodes, the IR-RS mechanism of NiO nanodots is attributed to the built-in isotype homojunction that is modulated by oxygen migration, which is evidenced by the contrast experiments in air and in vaccum. Under the worst-case scenario, the maximum crossbar array size is calculated to be as large as 3 Mbit with 10% read margin, outperforming most cases of the interface-dominated IR-RS reported previously. This is the first work reporting the IR-RS dominated by the built-in homojunction, which was studied on the nanoscale, and its superiority for selection device-free memory application has also been demonstrated. To this end, NiO is suggested as a promising material for memory applications, combining with its other advantages.

## 4. Experimental Section

*Fabrication of NiO nanodots*: Purchased AAO templates were transferred to commercial Pt/Ti/SiO$_2$/Si substrates. Using a sintered ceramic NiO target, NiO nanodots were prepared by PLD method, which was equipped with a KrF excimer laser ($\lambda$ = 248 nm, pulse energy = 110 mJ, frequency = 3 Hz, Lambda Physik). The deposition temperature was 500 ℃. The base pressure was $1 \times 10^{-3}$ Pa, and the oxygen pressure during deposition was 1.5 Pa. For the Au/NiO nanodot/Pt samples, the Au top electrodes were deposited by magnetron sputtering after the deposition of NiO. Finally, the AAO templates were lifted-off by mechanical method.

*Characterizations*: The surface topography of NiO nanodots were characterized by the AFM mode of a Bruker Multi-mode 8 SPM. The SEM image was obtained in an S-4800 SEM operated at an acceleration voltage of 5 kV (Hitachi, Japan). Samples for cross-section TEM were prepared using a standard procedure consisting of gluing, cutting, mechanical polishing, dimpling, and ion milling. The TEM measurements were operated on a Tecnai F20 transmission electron microscope operated at an acceleration voltage of 200 kV (FEI, USA).

*Electrical measurements*: The electric measurements were all accomplished by CAFM setups, including a Bruker Multi-mode 8 SPM with a CAFM module, whose compliance current is 12.28 nA, and a Seiko SII E-SWEEP AFM with a current collector, whose compliance current is 100 pA, the former enables only measurement in atmosphere, while the latter can perform measurements in vacuum. When a different voltage-sweep sequence was needed, an external Keithley 2400 source meter was employed, which also enables different compliance currents.

**Acknowledgements**
We gratefully acknowledge Y. Wang, X. Shi and X. Lai for help with the CAFM measurements. This work was supported by the National Science Foundation of China (Grant Nos. 11134007, 51572150)), and 973 project of the Ministry of Science and Technology of China (Grant No. 2015CB921402).

Received: ((will be filled in by the editorial staff))
Revised: ((will be filled in by the editorial staff))
Published online: ((will be filled in by the editorial staff))

[1] H. Akinaga, H. Shima, *Proc. IEEE* **2010**, *98*, 2237.

[2] M. Lubben, P. Karakolis, V. Ioannou-Sougleridis, P. Normand, P. Dimitrakis, I. Valov, *Adv. Mater.* **2015**, *27*, 6202.

[3] T. -W. Kim, H. Choi, S. -H. Oh, G. Wang, D. -Y. Kim, H. Hwang, T. Lee, *Adv. Mater.* **2009**, *21*, 2497.

[4] M. -J. Lee, S. I. Kim, C. B. Lee, H. Yin, S. -E. Ahn, B. S. Kang, K. H. Kim, J. C. Park, C. J. Kim, I. Song, S. W. Kim, G. Stefanovich, J. H. Lee, S. J. Chung, Y. H. Kim, Y. Park, *Adv. Funct. Mater.* **2009**, *19*, 1587.

[5] M. -J. Lee, S. Seo, D. -C. Kim, C. S. -E. Ahn, D. H. Seo, I. -K. Yoo, I. -G. Baek, D. -S. Kim, I. -S. Byun, S. -H. Kim, I. -R. Hwang, J. -S. Kim, S. -H. Jeon, B. H. Park, *Adv. Mater.* **2007**, *19*, 73.

[6] G. Wang, A. C. Lanchner, J. Lin, D. Natelson, K. V. Palem, J. M. Tour, *Adv. Mater.* **2013**, *25*, 4789.

[7] M. -J. Lee, Y. Park, D. -S. Suh, E. -H. Lee, S. Seo, D. -C. Kim, R. Jung, B. -S. Kang, S. -E. Ahn, C. B. Lee, D. H. Seo, Y. -K. Cha, I. -K. Yoo, J. -S. Kim, B. H. Park, *Adv. Mater.* **2007**, *19*, 3919.

[8] S. H. Chang, S. B. Lee, D. Y. Jeon, S. J. Park, G. T. Kim, S. M. Yang, S. C. Chae, H. K. Yoo, B. S. Kang, M. -J. Lee, T. W. Noh, *Adv. Mater.* **2011**, *23*, 4063.

[9] C. Chen, F. Pan, Z. S. Wang, J. Yang, F. Zeng, *J. Appl. Phys.* **2012**, *111*, 013702.

[10] J. Y. Kwon, J. H. Park, T. G. Kim, *Appl. Phys. Lett.* **2015**, *106*, 223506.

[11] E. Linn, R. Rosezin, C. Kugeler, R. Waser, *Nat. Mater.* **2010**, *9*, 403.

[12] Y. C. Bae, A. R. Lee, J. B. Lee, J. H. Koo, K. C. Kwon, J. G. Park, H. S. Im, J. P. Hong, *Adv. Funct. Mater.* **2012**, *22*, 709.

[13] S. Gao, F. Zeng, F. Li, M. Wang, H. Mao, G. Wang, C. Song, F. Pan, *Nanoscale* **2015**, *7*, 6031.

[14] Y. Dong, G. Yu, M. C. MaAlpine, W. Lu, C. M. Lieber, *Nano Lett.* **2008**, *8*, 386.

[15] A. Sawa, T. Fujii, M. Kawasaki, Y. Tokura, *Appl. Phys. Lett.* **2004**, *85*, 4073.

[16] Q. Zuo, S. Long, Q. Liu, S. Zhang, Q. Wang, Y. Li, Y. Wang, M. Liu, *J. Appl. Phys.* **2009**, *106*, 073724.

[17] X. A. Tran, B. Gao, J. F. Kang, X. Wu, L. Wu, Z. Fang, Z. R. Wang, K. L. Pey, Y. C. Yeo, A.Y. Du, M. Liu, B. Y. Nguyen, M. F. Li, H. Y. Yu, *IEDM Tech. Dig.* **2011**, 31.2.

[18] J. H. Yoon, S. J. Song, I. -H. Yoo, J. Y. Seok, K. J. Yoon, D. E. Kwon, T. H. Park, C. S. Hwang, *Adv. Funct. Mater.* **2014**, *24*, 5086.

[19] H. Shima, N. Zhong, H. Akinaga, *Appl. Phys. Lett.* **2009**, *94*, 082905.

[20] G. Wang, J. -H. Lee, Y. Yang, G. Ruan, N. D. Kim, Y. Ji, J. M. Tour, *Nano Lett.* 2**015**, *15*, 6009.

[21] C. H. Nieh, M. L. Lu, T. M. Weng, Y. F. Chen, *Appl. Phys. Lett.* **2014**, *104*, 213501.

[22] R. Meyer, L. Schloss, J. Brewer, R. Lambertson, W. Kinney, J. Sanchez, D. Rinnerson, *Non-Volatile Memory Technology Symp. 2008*, **2008**, 1.

[23] R. Meyer, R. Liedtke, R. Waser, *Appl. Phys. Lett.* **2005**, *86*, 112904.

[24] D. B. Strukov, J. L. Borghetti, R. S. Williams, *Small* **2009**, *9*, 1058.

[25] S. M. Sze, K. N. Kwok, *Physics of Semiconductor Devices*, 3rd ed.; John Wiley & Sons, New York, **2007**.

[26] S. Seo, M. J. Lee, D. C. Kim, S. E. Ahn, B. -H. Park, Y. S. Kim, I. K. Yoo, I. S. Byun, I. R. Hwang, S. H. Kim, J. -S. Kim, J. S. Choi, J. H. Lee, S. H. Jeon, S. H. Hong, B. H. Park, *Appl. Phys. Lett.* **2005**, *87*, 263507.

[27] K. Oka, T. Yanagida, K. Nagashima, M. Kanai, B. Xu, B. H. Park, H. Katayama-Yoshida, T. Kawai, *J. Am. Chem. Soc.* **2012**, *134*, 2535.

[28] Z. Sun, Y. Zhao, M. He, L. Gu, C. Ma, K. Jin, D. Zhao, N. Luo, Q. Zhang, N. Wang, W. Duan, C. -W. Nan, *ACS Appl. Mater. Interfaces* **2016**, *8*, 11583.

[29] T. Yanagida, K. Nagashima, K. Oka, M. Kanai, A. Klamchuen, B. H. Park, T. Kawai, *Sci. Rep.* **2013**, *3*, 1657.

[30] W. Lee, H. Han, A. Lotnyk, M. A. Schubert, S. Senz, M. Alexe, D. Hesse, S. Baik, U. Gosele, *Nat. Nanotechnol.* **2008**, *3*, 402.

[31] J. Y. Son, Y. -H. Shin, H. Kim, H. M. Jang, *ACS Nano* **2010**, *4*, 2655.

[32] S. Hong, T. Choi, J. H. Jeon, Y. Kim, H. Lee, H. -Y. Joo, I. Hwang, J. -S. Kim, S. -O. Kang, S. V. Kalinin, B. H. Park, *Adv. Mater.* **2013**, *25*, 2339.

[33] C. Park, S. H. Jeon, S. C. Chae, S. Han, B. H. Park, S. Seo, D. -W. Kim, *Appl. Phys. Lett.* **2008**, *93*, 042102.

[34] J. J. Yang, M. D. Pickett, X. Li, D. A. A. Ohlberg, D. R. Stewart, R. S. Williams, *Nat. Nanotechnol.* **2008**, *3*, 429.

[35] S. Menzel, M. Waters, A. Marchewka, U. Bottger, R. Dittmann, R. Waser, *Adv. Funct. Mater.* **2011**, *21*, 4487.

[36] Y. Kim, S. J. Kelly, A. Morozovska, E. K. Rahani, E. Strelcov, E. Eliseev, S. Jesse, M. D. Biegalski, N. Balke, N. Benedek, D. Strukov, J. Aarts, I. Hwang, S. Oh, J. S. Choi, T. Choi, B. H. Park, V. B. Shenoy, P. Maksymovych, S. V. Kalinin, *Nano Lett.* **2013**, *13*, 4068.

[37] H. Lee, H. Kim, T. N. Van, D. -W. Kim, J. Y. Park, *ACS Appl. Mater. Interfaces* **2013**, *5*, 11668.

[38] A. Siemon, T. Breuer, N. Aslam, S. Ferch, W. Kim, J. Van den Hurk, V. Rana, S. Hoffman-Eifert, R. Waser, S. Menzel, E. Linn, *Adv. Funct. Mater.* **2015**, *25*, 6414.

[39] J. Y. Seok, S. J. Song, J. H. Yoon, K. J. Yoon, T. H. Park, D. E. Kwon, H. Lim, G. H. Kim, D. S. Jeong, C. S. Hwang, *Adv. Funct. Mater.* **2014**, *23*, 5316.

[40] J. H. Jeon, H. -Y. Joo, Y. -M. Kim, D. H. Lee, J. -S. Kim, Y. S. Kim, T. Choi, B. H. Park, *Sci. Rep.* **2016**, *6*, 23299.

[41] C. Yoshida, K. Kinoshita, T. Yamasaki, Y. Sugiyama, *Appl. Phys. Lett.* **2008**, *93*, 042106.

[42] M. H. Lee, S. J. Song, K. M. Kim, G. H. Kim, J. Y. Seok, J. H. Yoon, C. S. Hwang, *Appl. Phys. Lett.* **2010**, *97*, 062909.

[43] M. D. Irwin, D. B. Buchholz, A. W. Hains, R. P. H. Chang, T. J. Marks, *Proc. Nat'l Acad. Soc. USA* **2008**, *105*, 2783.

[44] H. B. Michaelson, *J. Appl. Phys.* **1977**, *48*, 4729.

[45] C. M. Osburn, R. W. Vest, *J. Phys. Chem. Solids* **1971**, *32*, 1331.

[46] C. M. Osburn, R. W. Vest, *J. Phys. Chem. Solids* **1971**, *32*, 1343.

[47] R. Haugsrud, T. Norby, *Solid State Ionics* **1998**, *111*, 323.

[48] R. Molaei, R. Bayati, J. Narayan, *Cryst. Growth Des.* **2013**, *13*, 5459.

[49] T. -G. Seong, M. -R. Joung, J. -W. Sun, M. K. Yang, J. -K. Lee, J. W. Moon, J. Roh, S. Nahm, *Jpn. J. Appl. Phys.* **2012**, *51*, 041102.

[50] M. T. Greiner, L. Chai, M. G. Helander, W. -M. Tang, Z. -H. Lu, *Adv. Funct. Mater.* **2012**, *22*, 4557.

[51] K. R. Virwani, A. P. Malshe, J. P. Rajurkar, *Phys. Rev. Lett.* **2007**, *99*, 017601.

[52] R. Molaei, R. Bayati, J. Narayan, *Cryst. Growth Des.* **2013**, *13*, 5459.

[53] T. Tsuruoka, K. Terabe, T. Hasegawa, I. Valov, R. Waser, M. Aono, *Adv. Funct. Mater.* **2012**, *22*, 70.

[54] S. Tappertzhofen, I. Valov, T. Tsuruoka, T. Hasegawa, R. Waser, M. Aono, *ACS Nano* **2013**, *7*, 6396.

[55] F. Messerschmitt, M. Kubicek, J. L. M. Rupp, *Adv. Funct. Mater.* **2015**, *25*, 5117.

[56] R. Waser, T. Baiatu, K. -H.Hardtl, *J. Am. Ceram. Soc.* **1990**, *73*, 1654.

[57] C.-H. Yang, J. Seidel, S. Y. Kim, P. B. Rossen, P. Yu, M. Gajek, Y. H. Chu, L. W. Martin, M. B. Holcomb, Q. He, P. Maksymovych, N. Balke, S. V. Kalinin, A. P. Baddorf, S. R. Basu, M. L. Scullin, R. Ramesh, *Nat. Mater.* **2009**, *8*, 485.

[58] V. Garcia, S. Fusil, K. Bouzehouane, S. Enouz-Vedrenne, N. D. Mathur, A. Barthelemy, M. Bibes, *Nature* **2009**, *460*, 81.

[59] V. Garcia, M. Bibes, *Nat. Commun.* **2014**, *5*, 4289.

[60] G. Catalan, J. Seidel, R. Ramesh, J. F. Scott, *Rev. Mod. Phys.* **2012**, *84*, 119.

[61] K. Szot, W. Speier, G. Bihlmayer, R. Waser, *Nat. Mater.* **2006**, *5*, 312.

[62] J. Qi, M. Olmedo, J. Ren, N. Zhan, J. Zhao, J. -G. Zheng, J. Liu, *ACS Nano* **2012**, *6*, 1051.

[63] M. Sowinska, T. Bertaud, D. Walczyk, S. Thiess, M. A. Schubert, M. Lukosius, W. Drube, Ch. Walczyk, T. Schroeder, *Appl. Phys. Lett.* **2012**, *100*, 233509.

[64] Q. Zuo, S. Long, S. Yang, Q. Liu, L. Shao, Q. Wang, S. Zhang, Y. Li, Y. Wang, M. Liu, *IEEE Electron Device Lett.* **2010**, *31*, 344.

[65] A. Flocke, T. G. Noll, in *Proc. 33rd European Solid State Circuits Conf.* **2007**, 328.

[66] M. -J. Lee, S. -E. Ahn, C. B. Lee, C. -J. Kim, S. Jeon, U. -I. Chung, I. -K. Yoo, G.-S. Park, S. Han, I. R. Hwang, B. -H. Park, *ACS Appl. Mater. Interfaces* **2011**, *3*, 4475.

**TOC Entry:**

**NiO nanodots fabricated with ultrathin anodic aluminum oxide (AAO) templates** show typical intrinsically rectifying-resistive switching behaviors, which are exclusively attributed to the built-in isotype homojunction induced by oxygen migration. Under the worst-case scenario, the maximum crossbar array size is calculated to be as large as 3 Mbit. This work has demonstrated the superiority of NiO for memory applications.

**Keywords:** memristor, anodic aluminum oxide (AAO), NiO nanodots, intrinsically rectifying-resistive switching, built-in homojunction

By *Zhong Sun*, *Linlin Wei*, *Ce Feng*, *Peixian Miao*, *Meiqi Guo*, *Huaixin Yang*, *Jianqi Li*, *Yonggang Zhao*[*]

**Title:** Built-in Homojunction Dominated Intrinsically Rectifying-Resistive Switching in NiO Nanodots for Selection Device-Free Memory Application

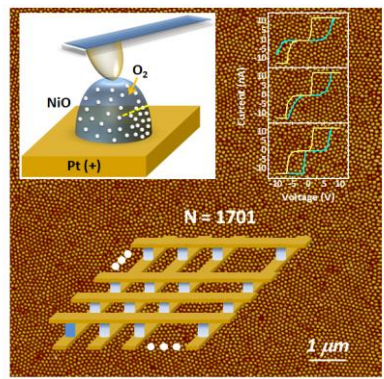